\documentclass[aip,jmp,amsmath,amssymb,reprint,onecolumn]{revtex4-1}

\usepackage{dcolumn}
\usepackage{bm}
\usepackage{amsfonts}
\usepackage{euscript}
\usepackage{newlfont}
\usepackage{euscript}
\usepackage{slashed}
\usepackage{url}
\newtheorem{theorem}{Theorem}

\newtheorem{conjecture}{Conjecture}
\newtheorem{definition}[theorem]{Definition}

\renewcommand{\div}{\operatorname{div}}
\newcommand{\tr}{\operatorname{Tr}}
\newcommand{\R}{\mathbb{R}}
\newcommand{\st}{\colon\>}
\newcommand{\C}{\mathbb{C}}
\newcommand{\End}{\operatorname{End}}

\newcommand{\gmp}{g_{\text{MP}}}
\newcommand{\pmp}{\phi_{\text{MP}}}
\newcommand{\gb}{\overline g}
\newcommand{\Mb}{\overline M}

\newcommand{\dirac}{\slashed{D}}

\begin{document}


\title{Rigidity in the Positive Mass Theorem with Charge}

\author{Marcus A.~Khuri}
 \affiliation{Department of Mathematics, Stony Brook University, Stony Brook, NY 11794}
 \email{khuri@math.sunysb.edu}
 \thanks{The first author acknowledges the support of NSF Grants DMS-1007156, DMS-1308753, and a Sloan Research Fellowship.}
\author{Gilbert Weinstein}
 \email{gilbert.weinstein@gmail.com}
 \affiliation{Physics Dept.\ and Dept.\ of  Computer Science and Mathematics, Ariel University of Samaria, Ariel
40700, Israel}

\date{\today}

\begin{abstract}
In this paper we show how a natural coupling of the Dirac equation with the generalized Jang equation,
leads to a proof of the rigidity statement in the positive mass theorem with charge, without the maximal slicing
condition, provided a solution to the coupled system exists.
\end{abstract}


\maketitle

\section{Introduction}
\label{sec1} \setcounter{equation}{0}
\setcounter{section}{1}

Consider an initial data set for the Einstein equations $(M,g,k)$, where $M$ is a
3-manifold, $g$ a Riemannian metric, and $k$ a symmetric 2-tensor.  The
energy and momentum densities are given by
\begin{align*}
	2\mu &= R+(\tr k)^{2}-|k|^{2}, \\
	\quad J &= \div(k- (\tr k) g),
\end{align*}
and the dominant energy condition is
\begin{equation}	\label{eq:dec}
      \mu \geq |J|.
\end{equation}

For simplicity, we will throughout assume that the data set possesses a single strongly asymptotically flat end,
that is, a region of $M$ which is diffeomorphic to the
complement of a ball in $\R^3$, with the property that the metric $g$ and tensor $k$ expressed in coordinates
induced by this
diffeomorphism decay with the following rates:
\[
    g_{ij} - \delta_{ij} = O_2(r^{-1}), \quad k_{ij} = o_1(r^{-2}),
\]
where $r$ is the Euclidean distance in these asymptotically flat coordinates, and where $O_p
(r^{-\beta})$
($o_p(r^{-\beta})$ respectively) means a function $f$ such that $\sum_{j=0}^p r^{\beta+j} |D^j
f|$ is
bounded (tends to zero respectively) as $r\to\infty$. We will say that the initial data set is strongly asymptotically flat if it is the union of a compact set
and a strongly asymptotically flat end.
The {total mass} of the end is defined by
\[
  m = \frac1{16\pi} \lim_{r\to\infty} \int_{S_r} (g_{ij,j} - g_{jj,i}) \nu^i,
\]
where $S_r$ is the coordinate sphere of radius $r$ and $\nu$ is its unit outward normal.

The Riemannian manifold $(M,g)$ is assumed to be complete with boundary, where the boundary $\partial M$ is an outermost apparent
horizon. By an apparent horizon, we mean that $\partial M$ is compact and each component has mean curvature
$H_{\partial M}=\pm\tr_{\partial M} k$ where the trace $\tr_{\partial M}$ is taken with respect to the induced metric on
$\partial M$. The apparent horizon is outermost if it is not `contained'
in any other apparent horizon. To make sense of the term `contain', we restrict our attention
to surfaces which bound a $3$-dimensional region containing the boundary. Recall the positive mass theorem.

\begin{theorem}	\label{thm:pmt}
Let $(M,g,k)$ be initial data with a strongly asymptotically flat end and satisfying the dominant energy
condition. Then
\[
  m\geq 0,
\]
and equality holds if and only if $(M,g,k)$ arises from Minkowski space.
\end{theorem}

The case of equality in this theorem is referred to as the rigidity statement. Note that our hypothesis of strong
asymptotic flatness implies the vanishing of the ADM linear momentum. Under weaker hypotheses, in which the linear
momentum is possibly non-zero, the rigidity statement was proved in Refs.~\onlinecite{yip1987,beigchrusciel1996}.

Two entirely different approaches to prove this theorem are relevant to our work in this
paper.  The first one, due to Schoen and Yau was given in two installments, first in the maximal
slice case, that is when $\tr k=0$, using minimal surface theory \cite{schoenyau1979}.  This result was
then used to prove the general
case by applying the Jang equation \cite{schoenyau81}.  The main idea in this second step is to look for
a hypersurface $\overline M$, in the Riemannian product $(M \times \R,g+dt^2)$, given as the graph of
a function $f$ on $M$, and such that its mean curvature satisfies $H_{\overline M} = \tr_{\overline M} k$.
This last equation is known as the Jang equation, and its validity ensures that the
the scalar curvature of $\overline g=g+df^2$ (the induced metric on $\overline M$), is in some sense weakly nonnegative,
so that there is a conformal metric
$\tilde g = u^4\overline{g}$ with zero scalar curvature. One may then apply the
maximal slice result to obtain the desired conclusion.

A second approach is due to
Witten and is based on solutions of the Dirac equation \cite{witten1981}.  Here, using a harmonic spinor $\psi$, which
tends to a constant spinor $\psi_0$ at infinity with $|\psi_0|\to 1$, one derives an integral identity which exhibits
the mass explicitly as the
integral of a nonnegative
quantity over $M$ \cite{leeparker1987}.

Now consider the generalization of Theorem~\ref{thm:pmt} to initial data sets for the Einstein-Maxwell equations.
For simplicity, we restrict ourselves to the case where the initial magnetic field vanishes.  Thus,
we consider initial data sets
$(M,g,k,E)$ where $E$ is a vector field on $M$.  We say that this data set is strongly
asymptotically flat if in addition to the decay condition above we also have
\[
  E = O_1(r^{-2}).
\]
The energy density of the matter content after the contribution of the electric field has been removed is given by
\[
  	2\mu_0 = R+(\tr k)^{2}-|k|^{2} - 2|E|^2.
\]
Define half the charge density by
\[
    \rho = \frac{1}{2}\div E,
\]
and denote the total charge
\[
  Q = \lim_{r\to\infty} \frac1{4\pi} \int_{S_r} g(E,\nu).
\]
The charged dominant energy condition, in Gaussian units with $c=G=1$, then takes the form
\begin{equation}	\label{eq:doc}
   \mu_0 \geq |J| + |\rho|.
\end{equation}

For the rigidity statement in the positive mass theorem to follow, the role of Minkowski space is replaced by the
Majumdar-Papapetrou spacetime, which we will denote by MP. This spacetime has the manifold structure $\R\times
(\R^3\setminus\{p_1,\dots,p_l\})$ and is equipped with the metric
\[
  \gmp = \pmp^2\, dt^2 + \pmp^{-2}\delta,
\]
where $p_1,\dots,p_l$ are $l$ points in $\R^3$,
\begin{equation}	\label{pmp}
  \pmp = \left(1 + \sum_{i=1}^l \frac {m_i}{r_i} \right)^{-1},
\end{equation}
$\delta$ is the flat metric on $\R^3$, $r_i$ is the Euclidean distance to $p_i$, and the constants $m_{i}$ are all
nonnegative.  With the
electric field
$E_{\text{MP}}=\nabla\log\pmp$, this spacetime can be interpreted as $l$ black holes in
equilibrium, where
the $i$-th black hole
has mass $m_i$ and charge $m_i$. In other words, the Lorentzian spacetime $(\R\times
(\R^3\setminus\{p_1,\dots,p_l\}),\gmp)$ is static and electrovacuum,
and asymptocially flat except for $l$ cylindrical ends.

The first part of the following theorem was proved in Ref.~\onlinecite{gibbonshawkinghorowitzperry} (see also
Ref.~\onlinecite{bartnikchrusciel2005}), using a modification of Witten's approach. The rigidity statement was
established in Ref.~\onlinecite{chruscielrealltod2006}.

\begin{theorem} \label{thm:cpmt}
\hspace*{-.2ex}Let $(M,g,k,E)$ be initial data for the Einstein-Maxwell
equations with a strongly asymptotically flat end, and satisfying the charged dominant energy condition, then
\[
  m\geq |Q|.
\]
Suppose in addition that $\tr k=0$, then equality holds if and only if the data set $(M,g,k,E)$ arises from the
Majumdar-Papapetrou spacetime.
\end{theorem}

The purpose of the present paper is to remove the maximal slice hypothesis in the above theorem, by coupling the
Dirac equation with a natural modification of the Jang equation.  This latter equation was introduced by Bray and Khuri
\cite{braykhuri2010, braykhuri2011} in an attempt to reduce the
general Penrose inequality to the Riemannian Penrose inequality. This was carried out
successfully in the spherically symmetric case. In this approach, one still
looks for a hypersurface $\Mb$ satisfying an equation of the form $H_{\overline{M}}=\tr_{\Mb} K$ (where $K$ is an
extension of $k$)
but in a warped product metric $g+\phi^{2}dt^{2}$
rather than in
the product metric.  This
is of course better adapted to the task at hand, since the MP metric is itself a warped
product metric.

In the next section, we will formulate the coupled Dirac-Jang system of equations.  In
Section~\ref{sec:cpmt}, it
will shown that if this system has a solution, then the rigidity statement of Theorem~\ref{thm:cpmt} follows without
the maximal slice
hypothesis. Finally in Section~\ref{sec:analysis}, we present evidence which indicates that the
system of Section~\ref{sec:system} can be solved.  Some of the most technical computations are
relegated to the appendices.

\section{The coupled system}	\label{sec:system}

\subsection{The Generalized Jang Equation}

Let $\phi\colon M\to\R$ be a positive function, and consider the warped product metric $g+\phi^2 dt^2$ on
$M\times\mathbb{R}$.
Let $f\colon M\to\R$ be a smooth function, and denote its graph by $\Mb=\{(x,f(x))\st x\in M\} \subset M\times\R$. The
induced metric on $\overline{M}$ arising from the warped product metric is given by $\gb=g+\phi^{2}df^{2}$. The
generalized Jang equation is then given by
$H_{\overline{M}}=\tr_{\overline{M}}K$, where $H_{\overline{M}}$ is the mean curvature of the graph and $K$ is a
particular extension of the
initial data $k$ to the 4-manifold $M\times\mathbb{R}$ (see Ref.~\onlinecite{braykhuri2010}). In local coordinates this
equation
becomes
\begin{equation}	\label{eq:mje}
  \left( g^{ij}-\frac{\phi^{2}f^{i}f^{j}}{1+\phi^{2} |\nabla f|^{2}}\right)
  \left( \frac{\phi \nabla_{ij}f+\phi_{i}f_{j}+\phi_{j}f_{i}} {\sqrt{1+\phi^{2} |\nabla
f|^{2}}}-k_{ij} \right)
  =0,
\end{equation}
where $\nabla_{ij}f$ are covariant derivatives with respect to $g$ and $f^{i}=g^{ij}f_{j}$. The purpose of the generalized
Jang equation is
to give positivity properties to the scalar curvature of $\overline{g}$; this will be discussed further below. The basic
existence theory
for this equation has been established in Ref.~\onlinecite{hankhuri2012}.

\subsection{The Dirac Equation}

Dirac spinors are cross sections of a vector bundle $\mathcal{S}$ associated to a principal bundle over
$\Mb$ with structure group $SL(2,\C)$, the simply connected double cover of $SO(3,1)$,
see Ref.~\onlinecite{parkertaubes1982} for details. We write a Dirac equation on $\Mb$ which will then
couple with the generalized Jang equation. The vector bundle $\mathcal{S}$ is
equipped with a connection compatible with the Jang metric $\gb$ as follows
\[
 \overline{\nabla}_{e_{i}}=e_{i}+\frac{1}{4}\, \overline{\omega}_{ijl}\,
 c(e^{j})\, c(e^{l} )
\]
where $\overline{\omega}_{ijl}$ are the affine connection coefficients of the Jang
metric $\gb$ defined by
\[
 \gb(\overline{\nabla}_{e_i} e_j,e_l) = \overline{\omega}_{ijl},
\]
$e_1,e_2,e_3$ is an orthonormal frame on $\Mb$, and $c\colon T^* \Mb\to \End(\mathcal{S})$ is global
Clifford multiplication.
Consider the Einstein-Maxwell spin connection on $\mathcal{S}$
\begin{equation} \label{connection}
   \nabla_{e_{i}}= \overline{\nabla}_{e_i}
   -\frac{1}{2} \, c(\overline{E})\, c(e_{i})\, c(e_{0}),
\end{equation}
where $\overline{E}$ is the electric field associated with the Jang surface to be defined
below, and $e_0$ is the unit normal to $\Mb$.
An important observation is that because of the electric field contribution this connection
is \textit{not} metric compatible.

In order to define $\overline{E}$, let $F=F_{ab}\,dx^a\wedge dx^b$ be the field strength tensor on $M\times\mathbb{R}$,
given by $F_{0i}=\phi E_{i}$ and
$F_{ij}=0$ for $i,j=1,2,3$, where $x^{0}=t$, and the remaining $x^{i}$, $i=1,2,3$, are local coordinates on
$M$. Then set
\begin{equation*}
  \overline{E}_{i}=F(N,X_{i})=\frac{E_i + \phi^2 f_i f^j E_j}{\sqrt{1 + \phi^2 |\nabla f|^2_g}},
\end{equation*}
where $X_{i}=\partial_{i}+f_{i}\partial_{t}$, $i=1,2,3$ are basis elements for the tangent space to
$\Mb$ and
\begin{equation*}
  N=\frac{f^{i}\partial_{i}-\phi^{-2}\partial_{t}}{\sqrt{\phi^{-2}+|\nabla f|^{2}}}
\end{equation*}
is the unit normal to $\Mb$. This induced electric field on the Jang surface first appeared in
Ref.~\onlinecite{disconzikhuri}, and
has special
properties to be exploited below.

Let $\Gamma(\mathcal{S})$ be the space of cross-sections of the bundle $\mathcal{S}$. The Einstein-Maxwell Dirac
operator
$\dirac \colon
\Gamma(\mathcal{S})\to\Gamma(\mathcal{S})$ is now defined as usual
\begin{equation}	\label{op:dirac}
  \dirac\psi = \sum_{i=1}^3 c(e_i)\,
  \nabla_{e_i} \psi.
\end{equation}
A spinor $\psi$ on $\Mb$ is a harmonic spinor if it satisfies the
Dirac equation
\begin{equation}	\label{eq:dirac}
   \dirac \psi = 0.
\end{equation}

\subsection{The Dirac-Jang System}
We can now formulate the coupled system appropriate for our needs, by choosing the warping factor $\phi$ as follows.

\begin{definition}
Let $f\colon M\to\R$ with $f\in C^{2,\beta}(M)$, and $\psi\in\Gamma(S)$ with $\psi\in
C^{1,\beta}(M)$. We say that $(f,\psi)$ is a solution of the
Dirac-Jang system if $f$ satisfies~\eqref{eq:mje}, $\psi$ satisfies~\eqref{eq:dirac}, and
\begin{equation}	\label{eq:coupling}
  \phi=|\psi|^2.
\end{equation}
\end{definition}

Let us now investigate the appropriate asymptotics for solutions to the coupled Dirac-Jang system. First, the
asymptotics at
spatial infinity are standard and are given by
\begin{equation}\label{inftyasym}
f= O_{2}(|x|^{-\frac{1}{2}})\text{ }\text{ }\text{ and }\text{ }\text{ }\psi\rightarrow\psi_{0}\text{ }\text{ }\text{ as
}\text{ }\text{ }|x|\rightarrow\infty,
\end{equation}
where $\psi_{0}$ is a fixed constant spinor at infinity with $|\psi_{0}|=1$. Note that this fall-off for $f$ guarantees
that the mass of the Jang metric $\overline{g}$ agrees with that of $g$, and that the total charge of $\overline{E}$
agrees
with that of $E$.

In order to motivate the asymptotics at the apparent horizon, consider the MP spacetime with one
black hole (extreme Reissner-Nordstr\"{o}m spacetime) with metric
\begin{equation*}
-\left(1-\frac{m}{r}\right)^{2}dt^{2}+\left(1-\frac{m}{r}\right)^{-2}dr^{2}+r^{2}d\sigma^{2}.
\end{equation*}
Let $t=f(r)$ be a radial graph, with induced metric
\begin{equation*}
g=\left(\left(1-\frac{m}{r}\right)^{-2}-\left(1-\frac{m}{r}\right)^{2}f'^{2}\right)dr^{2}+r^{2}d\sigma^{2}.
\end{equation*}
As is calculated in Ref.~\onlinecite{braykhuri2011}, the second fundamental form of the graph is given by
\begin{equation*}
k_{ij}=\frac{\phi\nabla_{ij}f+\phi_{i}f_{j}+\phi_{j}f_{i}}{\sqrt{1+\phi^{2}|\nabla f|^{2}}}
\end{equation*}
where $\phi=1-\frac{m}{r}$ and the covariant derivatives are calculated with respect to the
metric $g$. Thus $(M=\mathbb{R}^{3}-B_{m}(0),g,k)$ forms an initial data set for which the graph $t=f(r)$ is a solution
of the generalized Jang
equation. If $f$ grows faster than $(r-m)^{-1}$ then $g$ will not be a Riemannian metric, that is, the graph will not be
spacelike.
Moreover if $f$ grows slower than $(r-m)^{-1}$ then $M$ will not have a boundary, but will rather have a cylindrical end
as in the MP initial data.
Thus, we will assume that
\begin{equation*}
f(r)\sim\int_{r}^{r_{0}}\left(1-\frac{m}{r}\right)^{-2}\sim(r-m)^{-1},
\end{equation*}
where $\sim$ indicates the presence of other added lower order terms. In order to see that the boundary is in fact a
future apparent horizon
with these asymptotics for $f$, let us note that
\begin{equation*}
g_{11}=\left(1-\frac{m}{r}\right)^{-2}-\left(1-\frac{m}{r}\right)^{2}f'^{2}\sim O(1).
\end{equation*}
Therefore the distance to the boundary is given by
\begin{equation*}
\tau=\int_{m}^{r}\sqrt{g_{11}}\sim r-m.
\end{equation*}
We now compute the future null expansion of the coordinate spheres with respect to the initial data metric $g$. A
standard formula yields
the mean curvature of coordinate spheres
\begin{equation*}
H_{S_{r}}=\frac{2\sqrt{g^{11}}}{r},
\end{equation*}
and the trace of the initial data $k$ over the coordinate spheres is given by
\begin{equation*}
\tr_{S_{r}}k=-\frac{\phi\gamma^{ij}\Gamma_{ij}^{1}f'}{\sqrt{1+\phi^{2}|\nabla f|^{2}}}=\frac{2}{r}\frac{\phi
g^{11}f'}{\sqrt{1+\phi^{2}g^{11}f'^{2}}},
\end{equation*}
where $\gamma_{ij}$ is the induced metric on $S_{r}$ and $\Gamma_{ij}^{l}$ are the Christoffel symbols of $g$.
It follows that the future null expansion becomes
\begin{align*}
\begin{split}
\theta_{+}=H_{S_{r}}+\tr_{S_{r}}k =& \frac{2}{r}\left(\sqrt{g^{11}}+\frac{\phi
g^{11}f'}{\sqrt{1+\phi^{2}g^{11}f'^{2}}}\right)\\
&=\frac{2}{r}\left(\frac{g^{11}}{1+\phi^{2}g^{11}f'^{2}}\right)\left(\sqrt{g^{11}}-\frac{\phi
g^{11}f'}{\sqrt{1+\phi^{2}g^{11}f'^{2}}}\right)^{-1}\\
&\sim (r-m)^{2}\\
&\sim \tau^{2},
\end{split}
\end{align*}
so that $\partial M$ is a future apparent horizon.

In conclusion, we expect that solutions of the Dirac-Jang system will follow similar asymptotics
\begin{equation}\label{bdryasym}
f\sim\tau^{-1}\text{ }\text{ }\text{ and }\text{ }\text{ }\psi\sim\tau^{\frac{1}{2}}\text{ }\text{ }\text{ as }\text{
}\text{ }
x\rightarrow\partial M.
\end{equation}
Notice also that with these asymptotics, the Jang metric still possesses an infinitely long cylindrical neck since
\begin{equation*}
\overline{g}=g+\phi^{2}df^{2}\sim O(1)+\tau^{2}\tau^{-4}\sim\tau^{-2},
\end{equation*}
and so
\begin{equation*}
\overline{\tau}=\int_{0}^{\tau}\sqrt{\overline{g}_{11}}\sim-\log\tau,
\end{equation*}
where $\overline{\tau}(x)=dist_{\overline{g}}(x,\partial M)$. Therefore the asymptotics lead to behavior for the Jang
metric which
is similar to that of initial data in the MP spacetime.

\section{The Positive mass theorem with charge} \label{sec:cpmt}

In this section we prove the following theorem.

\begin{theorem}	\label{cpmt}
\hspace*{-.2ex}Let $(M,g,k,E)$ be strongly asymptotically flat initial data for the Einstein-Maxwell equations
satisfying the charged
dominant energy condition, with total mass $m$ and total charge $Q$, and suppose
that the Dirac-Jang system has a solution $(f,\psi)$ which satisfies \eqref{inftyasym} and \eqref{bdryasym}.  Then
\begin{equation*}
m\geq |Q|,
\end{equation*}
and equality holds if and
only if $(M,g,k,E)$ arises from the Majumdar-Papapetrou spacetime.
\end{theorem}

Note that the $t=0$ slice of the Majumdar-Papapetrou spacetime does not fall under the hypotheses of this theorem, as it possesses cylindrical ends.
This is related to the fact that we employ blow-up solutions of the generalized Jang equation at the apparent horizon boundary. In order to treat
initial data containing cylindrical ends, more general boundary conditions for the generalized Jang equation should be utilized.

The starting point of the proof is the Lichnerowicz identity, which now takes the
form
\begin{equation}	\label{eq:lichnerowicz}
   \dirac^2 = \nabla^*\nabla + \frac14 \mathcal{R},
\end{equation}
where $\mathcal{R}$ is the following endomorphism of $\mathcal{S}$
\[
   \mathcal{R} = \overline{R} - 2|\overline{E}|^{2} - 2\overline\rho \, c(e_{0}),
\]
with $\overline{R}$ the scalar curvature of the Jang metric $\overline{g}$ and
$\overline\rho=\frac12\overline{\div}\!\text{ }\overline E$ the charge density on $\overline M$.
Applying~\eqref{eq:lichnerowicz} to the harmonic spinor
$\psi$, taking the inner product with $\psi$, and integrating by parts over $\Mb$ produces
\begin{equation}	\label{eq:int-identity}
   \int_{\Mb} |\nabla\psi|^2 + \frac14 \langle \mathcal{R}\psi, \psi\rangle = 4\pi(m-|Q|).
\end{equation}
The right hand side is the boundary term at spatial infinity, and no interior boundary terms appear in light of
the asymptotics \eqref{bdryasym}.

We will now show that the second term on the left hand side of \eqref{eq:int-identity} is nonnegative.
As proved in Ref.~\onlinecite{disconzikhuri}, our choice for the electric field $\overline
E$ on the Jang surface ensures that
\begin{gather}
  |E|\geq|\overline{E}|, \label{EvsEbar} \\
  \rho=\overline\rho \, \sqrt{1+\phi^{2}|\nabla f|^{2}}. \label{rho}
\end{gather}
Moreover, due to the fall off rate of the Jang graph the total charge is unchanged
\begin{equation}
  \lim_{r\to\infty}\frac1{4\pi}\int_{S_r} \gb(\overline E,\overline \nu) = Q.
\end{equation}

In Ref.~\onlinecite{braykhuri2010}, it was shown that the generalized Jang equation implies the following formula for the scalar
curvature of $\overline{g}$
\begin{equation} \label{Rbar}
  \overline{R} =
  2(\mu-J(w))
  +|h-K|^{2}+2|q|^{2}-2\phi^{-1}\, \overline{\div}(\phi q),
\end{equation}
where $h$ is the second fundamental form
of the Jang surface $\Mb$, $K$ is a specially chosen extension of the initial data $k$, and
\begin{gather}
  w^{i}=\frac{\phi f^{i}}{\sqrt{1+\phi^{2}|\nabla f|^{2}}}, \label{eq:w} \\[1ex]
  q_{i}=w^{j}(h_{ij}-K_{ij}). \label{eq:q}
\end{gather}
Notice that \eqref{eq:doc}, \eqref{EvsEbar}, and \eqref{Rbar}, together with $|w|\leq1$
imply
\begin{equation}	\label{End}
\begin{aligned}
  \overline{R} - 2|\overline{E}|^{2}
  &= 2(\mu_0-J(w))+2(|E|^{2}-|\overline{E}|^{2})
    +|h-K|^{2}+2|q|^{2}-2\phi^{-1}\overline{\div}(\phi q) \\
  &\geq 2|\rho| - 2\phi^{-1}\overline{\div} ( \phi q).
\end{aligned}
\end{equation}
We conclude, in view of the coupling $\phi=|\psi|^2$, and $|\rho|\geq|\overline\rho|$, that
the second term on the left hand side in~\eqref{eq:int-identity} satisfies
\begin{equation}	\label{2ndterm}
    \frac14 \langle \mathcal{R}\psi, \psi\rangle \geq \frac12 \left( |\rho|\, |\psi|^2 - \langle
     \overline\rho c(e_0)\psi, \psi \rangle -  \overline{\div}(\phi q) \right) \geq -
     \frac12 \overline{\div}(\phi q).
\end{equation}
Thus, \eqref{eq:int-identity} becomes
\begin{equation}	\label{mQ}
    \int_{\Mb}  |\nabla\psi|^2  = \int_{\Mb} |\nabla\psi|^2 -\frac12 \overline{\div}(\phi q)\leq 4\pi(m-|Q|).
\end{equation}
Note that the integral of the divergence term vanishes according to the asymptotics \eqref{inftyasym} and
\eqref{bdryasym}.

Now consider the case of equality in~\eqref{mQ}. It follows that the harmonic spinor
$\psi$ is covariantly constant $\nabla\psi=0$.
We now recall that $\nabla$ is not metric compatible, so that although $\psi$ is
parallel, $|\psi|$ is not constant.  In fact, we show in the Appendix~\ref{phi} that
$\phi=|\psi|^{2}$ satisfies the elliptic equation
\begin{equation}	\label{eq:phi}
  \overline{\Delta}\phi-\frac{1}{2}\overline{R}\phi=0,
\end{equation}
where $\overline{\Delta}$ is the Laplace-Beltrami operator with respect to $\overline{g}$.
Furthermore, it now follows from the energy condition~\eqref{eq:doc}, as well as \eqref{End} and \eqref{2ndterm},
that at each point of $M$ either
\begin{equation} \label{equality}
  \mu_0 = |J|
  = \rho = \overline\rho
  = |E|-|\overline{E}|
  =|h-K| = |q| = 0,
\end{equation}
or $|\psi|=0$. Since
\begin{equation}\label{bdryphi}
\phi(x)\rightarrow 0\text{ }\text{ }\text{ as }\text{ }\text{ }x\rightarrow\partial M,\text{ }\text{ }\text{ and }\text{
}\text{ }
\phi(x)\rightarrow 1\text{ }\text{ }\text{ as }\text{ }\text{ }x\rightarrow\infty,
\end{equation}
we may apply the Hopf maximum principle to conclude that $\phi>0$ on the interior of $M$. It then follows
that in fact \eqref{equality} holds without any preconditions.
In particular
\begin{equation}	\label{constraint}
  \overline{R} = 2|\overline{E}|^{2}.
\end{equation}
In conclusion, in view of~\eqref{constraint} and $\mu_0=|J|=\overline \rho=0$, we find that
$(\Mb,\overline{g},\overline{E})$ is a time symmetric electrovacuum
(and asymptotically flat) initial data set, with
\begin{equation*}
  m = |Q|.
\end{equation*}
According to Chrusciel, Reall, and Tod \cite{chruscielrealltod2006}, the only such electrovacuum
initial data is the MP initial data, and hence $\overline{g}=\gmp$ and $\overline{E}=E_{\text{MP}}$.
We now have $g=g_{MP}-\phi^{2}df^{2}$, so that the map $x\mapsto(x,f(x))$ yields an isometric
embedding of $(M,g)$ into the spacetime $(\R\times M,-\phi^{2}dt^{2}+\gmp)$. Next observe that
since $\pmp$ satisfies the same equation \eqref{eq:phi} and boundary conditions \eqref{bdryphi} as $\phi$,
we must have $\phi=\pmp$. Therefore $(\R\times M,-\phi^{2}dt^{2}+\gmp)$ is the MP spacetime.

It remains to show that $k$ and $E$ are respectively the second fundamental form of and induced electric field on the
isometric embedding $(M,g)\hookrightarrow MP$. This, however, follows from previous work. Namely,
since $|h-K|=0$ it is shown in Refs.~\onlinecite{braykhuri2010,braykhuri2011} that $k$ is the desired second fundamental
form,
and the fact that $E$ is the induced electric field on the embedding is shown in Ref.~\onlinecite{disconzikhuri}.

This completes the proof of Theorem~\ref{cpmt}.

\section{A conjecture and some evidence} \label{sec:analysis}

In this section, we present evidence for the following conjecture.

\begin{conjecture}
Let $(M,g,k,E)$ be a strongly asymptotically flat initial data set for the Einstein-Maxwell equations.
Then the Dirac-Jang system has a solution $(f,\psi)$ which satisfies \eqref{inftyasym} and \eqref{bdryasym}.
\end{conjecture}

We note that if proved, this statement, in conjunction with Theorem~\ref{cpmt}, would also
give a proof of the following conjecture.

\begin{conjecture}
Let $(M,g,k,E)$ be a strongly asymptotically flat initial data set for the Einstein-Maxwell equations satisfying the
charged dominant energy condition, with total mass $m$ and total charge $Q$.  Then
\begin{equation*}
m\geq |Q|,
\end{equation*}
and equality holds if and
only if $(M,g,k,E)$ arises from the Majumdar-Papapetrou spacetime.
\end{conjecture}

Observe that the Dirac-Jang system is truly a coupled system of equations, since the metric
appearing in the Dirac equation depends on the solution to the generalized Jang
equation. However both are elliptic, or degenerate elliptic in the
case of the generalized Jang equation (the degeneracy appears only at the
horizons), and so we have the Schauder estimates available to analyze
(heuristically) whether it is possible to solve this set of equations.
Ultimately we would like to apply a standard iteration procedure to
obtain existence. Namely, pick an arbitrary positive function $\phi_{0}$ with the
correct boundary conditions, and then use it to solve the generalized Jang
equation for $f_{0}$. Use $\phi_{0}$ and $f_{0}$ to construct
$\overline{g}_{0}$, and then solve the Dirac equation to obtain
$\psi_{1}$ and hence $\phi_{1}$. Continuing in this way, we obtain
sequences $\{\phi_{i}\}$ and $\{f_{i}\}$. The appropriate estimates
must be made if we are to show that this procedure converges. Below we will perform a
calculation which suggests that uniform estimates for the warping factor $\phi$ are possible.
Once this is accomplished, it is a relatively easy task to uniformly bound the corresponding solution
of the generalized Jang equation and the Dirac equation.

To begin, recall that the warping factor is set to be the norm squared of the Dirac spinor, that is $\phi=|\psi|^{2}$.
In the Appendix~\ref{phi} it is shown that $\phi$ solves an equation of the form
\begin{equation} \label{phi:rhs}
\overline{\Delta}\phi-\frac{1}{2}\overline{R}\phi=\mathcal{F},
\end{equation}
where $\mathcal{F}$ is a function of first derivatives of the spinor $\psi$ and first derivatives of $\overline{g}$.
The scalar curvature has a nice form given by~\eqref{End}.
In this formula
\begin{equation*}
h_{ij}=\frac{\phi\nabla_{ij}f+\phi_{i}f_{j}+\phi_{j}f_{i}+\phi^{2}\phi^{l}f_{l}f_{i}f_{j}}
{\sqrt{1+\phi^{2}|\nabla f|^{2}}}
\end{equation*}
is the second fundamental form of the Jang surface inside $(M\times\mathbb{R}, g+\phi^{2}dt^{2})$,
$q$ and $w$ are given by~\eqref{eq:q}--\eqref{eq:w}, and $K$ is the extension to $M\times\mathbb{R}$
of the initial
data $k$ given by
\begin{equation*}
K(\partial_{x^{i}},\partial_{x^{j}})=k_{ij},\text{ }\text{ }\text{ }
K(\partial_{x^{i}},\partial_{t})=K(\partial_{t},\partial_{x^{i}})=0,\text{ }\text{ }\text{ }
K(\partial_{t},\partial_{t})=\frac{\langle\phi\nabla f,\phi\nabla\phi\rangle}{\sqrt{1+\phi^{2}|\nabla f|^{2}}}.
\end{equation*}
We observe that the scalar curvature $\overline{R}$ contains two derivatives of $\phi$ and three derivatives
of $f$, and thus it appears that by applying the Schauder estimates to equation~\eqref{phi:rhs}, the best we could hope
for is an estimate of the form
\begin{equation} \label{estimate:f3}
|\phi|_{C^{2,\alpha}}\leq
C(|\phi|_{C^{2,\alpha}}+|f|_{C^{3,\alpha}}+|\psi|_{C^{1,\alpha}}+|\overline{g}|_{C^{1,\alpha}}),
\end{equation}
which of course is of no help at all. However, below, we shall calculate the divergence term in the expression for the
scalar curvature, and invoke the generalized Jang equation, to show that this simple estimate may be improved.

Let $\Gamma_{ij}^{l}$ and $\overline{\Gamma}_{ij}^{l}$ be the Christoffel symbols for the initial data metric $g$ and
the Jang metric $\overline{g}$, respectively. Also note that
\begin{equation*}
q_{j}=\frac{\phi f^{l}}{\sqrt{1+\phi^{2}|\nabla f|^{2}}}\left(\frac{\phi\nabla_{lj}f+\phi_{l}f_{j}+\phi_{j}f_{l}}
{\sqrt{1+\phi^{2}|\nabla f|^{2}}}-k_{ij}\right).
\end{equation*}
Then a straightforward calculation yields
\begin{align*}
\begin{split}
\overline{\mathrm{div}}(\phi q)=&\overline{g}^{ij}\overline{\nabla}_{i}(\phi q_{j})\\
=&\overline{g}^{ij}\phi_{i}q_{j}
+\frac{\phi\overline{g}^{ij}}{1+\phi^{2}|\nabla f|^{2}}
[2\phi\phi_{i}f^{l}\nabla_{lj}f+\phi^{2}\nabla_{li}f\nabla^{l}_{j}f
+\phi^{2}f^{l}\nabla_{i}\nabla_{l}\nabla_{j}f\\
&+\phi^{2}f^{l}\nabla_{lm}f(\Gamma_{ij}^{m}-\overline{\Gamma}_{ij}^{m})
+\phi f^{l}\phi_{l}\overline{\nabla}_{ij}f
+\phi f^{l}f_{j}\overline{\nabla}_{il}\phi+\phi\phi^{l}f_{j}\overline{\nabla}_{li}f\\
&+\phi_{i}\phi_{l}f^{l}f_{j}
+\phi_{i}\phi_{j}|\nabla f|^{2}+\phi|\nabla f|^{2}\overline{\nabla}_{ij}\phi
+2\phi f^{l}\phi_{j}\overline{\nabla}_{li}f\\
&-\phi f^{l}(1+\phi^{2}|\nabla f|^{2})^{-1}(\phi\nabla_{lj}f+\phi_{l}f_{j}+\phi_{j}f_{l})
(2\phi\phi_{i}|\nabla f|^{2}+2\phi^{2}f^{m}\nabla_{mi}f)]\\
&-\phi\overline{g}^{ij}\left(\frac{\phi f^{l}k_{lj}}{\sqrt{1+\phi^{2}|\nabla f|^{2}}}\right).
\end{split}
\end{align*}
Observe that (see Ref.~\onlinecite{braykhuri2011})
\begin{equation*}
\Gamma_{ij}^{m}-\overline{\Gamma}_{ij}^{m}=\phi\phi^{m}f_{i}f_{j}-\frac{\phi f^{m}h_{ij}}{\sqrt{1+\phi^{2}|\nabla
f|^{2}}},
\end{equation*}
and recall the Ricci commutation formula
\begin{equation*}
\nabla_{i}\nabla_{l}\nabla_{j}f=\nabla_{l}\nabla_{j}\nabla_{i}f-R_{mjil}f^{m}.
\end{equation*}
Therefore, with the aid of the generalized Jang equation we find
\begin{multline*}
\frac{\phi\overline{g}^{ij}}{\sqrt{1+\phi^{2}|\nabla f|^{2}}}\nabla_{l}\nabla_{j}\nabla_{i}f
=
\nabla_{l}\left(\frac{\phi\overline{g}^{ij}\nabla_{ij}f}{\sqrt{1+\phi^{2}|\nabla f|^{2}}}\right)
-\nabla_{ij}f\nabla_{l}\left(\frac{\overline{g}^{ij}\phi}{\sqrt{1+\phi^{2}|\nabla f|^{2}}}\right)\\
=
\nabla_{l}\left[\overline{g}^{ij}\left(k_{ij}-\frac{\phi_{i}f_{j}+\phi_{j}f_{i}}{\sqrt{1+\phi^{2}|\nabla
f|^{2}}}\right)\right]-\nabla_{ij}f\nabla_{l}\left(\frac{\overline{g}^{ij}\phi}{\sqrt{1+\phi^{2}|\nabla
f|^{2}}}\right)\\
=
-\nabla_{ij}f\nabla_{l}\left(\frac{\overline{g}^{ij}\phi}{\sqrt{1+\phi^{2}|\nabla f|^{2}}}\right)
+\nabla_{l}(\overline{g}^{ij}k_{ij})
-\nabla_{l}\left(\frac{\overline{g}^{ij}}{\sqrt{1+\phi^{2}|\nabla f|^{2}}}\right)(\phi_{i}f_{j}+\phi_{j}f_{i})\\
-\overline{g}^{ij}\left(\frac{\nabla_{li}\phi f_{j}+\phi_{i}\nabla_{lj}f+\nabla_{jl}\phi f_{i}
+\phi_{j}\nabla_{li}f}{\sqrt{1+\phi^{2}|\nabla f|^{2}}}\right).
\end{multline*}
It follows that
\begin{equation*}
\overline{\mathrm{div}}(\phi q)=\frac{\phi^{2}}{1+\phi^{2}|\nabla f|^{2}}\left[\left(
g^{ij}-\frac{\phi^{2}f^{i}f^{j}}{1+\phi^{2}|\nabla f|^{2}}\right)|\nabla f|^{2}
-\frac{f^{i}f^{j}}{1+\phi^{2}|\nabla f|^{2}}\right]\overline{\nabla}_{ij}\phi+\cdots,
\end{equation*}
where $\cdots$ represents terms depending only on first derivatives of $\phi$, first derivatives of $\overline{g}$,
and second derivatives of $f$.

This shows that $\overline{\mathrm{div}}(\phi q)$ is a degenerate elliptic operator for
$\phi$, since the coefficients of the principal symbol may be rewritten as
\begin{equation*}
g^{ij}|\nabla f|^{2}-f^{i}f^{j}\geq 0.
\end{equation*}
Therefore the equation~\eqref{phi:rhs} is actually a strictly elliptic operator for $\phi$, with coefficients
depending only on first derivatives of $\phi$, first derivatives of $\overline{g}$,
and second derivatives of $f$. We may now apply the Schauder estimates to obtain an improvement of
the estimate~\eqref{estimate:f3}:
\begin{equation*}
|\phi|_{C^{2,\alpha}}\leq
C(|\phi|_{C^{1,\alpha}}+|f|_{C^{2,\alpha}}+|\psi|_{C^{1,\alpha}}+|\overline{g}|_{C^{1,\alpha}}).
\end{equation*}
Moreover it is clear from the structure of the generalized Jang equation and the Dirac equation, that one should have
\begin{equation*}
|f|_{C^{2,\alpha}}\leq C(1+|\phi|_{C^{1,\alpha}}),
\end{equation*}
and
\begin{equation*}
|\psi|_{C^{1,\alpha}}\leq C|\overline{g}|_{C^{1,\alpha}}.
\end{equation*}
Since $\overline{g}=g+\phi^{2}df^{2}$, we also have
\begin{equation*}
|\overline{g}|_{C^{1,\alpha}}\leq C(1+|\phi|_{C^{1,\alpha}}+|f|_{C^{2,\alpha}})\leq C(1+|\phi|_{C^{1,\alpha}}).
\end{equation*}
Hence
\begin{equation*}
|\phi|_{C^{2,\alpha}}\leq C(1+|\phi|_{C^{1,\alpha}}).
\end{equation*}
These heuristic arguments yield strong evidence for a uniform bound on $\phi$, which as mentioned above, will lead
to uniform bounds on $f$ and $\psi$. Thus it appears that existence of a solution for the Dirac-Jang system is highly
likely, and rests on a uniform $C^{1,\alpha}$ bound for $\phi$.


\appendix

\section{The warping factor}	\label{phi}

Here we derive the equation~\eqref{phi:rhs} satisfied by the warping factor $\phi=|\psi|^{2}$.  Write
the connection~\eqref{connection} in the following way
\begin{equation*}
\nabla_{i}=\overline{\nabla}_{i}-A_{i},
\end{equation*}
where $\overline{\nabla}$ is the metric compatible (Levi-Civita) connection and
\begin{equation*}
A_{i}=\frac{1}{2}c(\overline{E})c(e_{i})c(e_{0}).
\end{equation*}
Direct computation yields
\begin{eqnarray*}
\overline{\Delta}|\psi|^{2}&=&\overline{\nabla}^{i}(\langle\overline{\nabla}_{i}\psi,\psi\rangle
+\langle\psi,\overline{\nabla}_{i}\psi\rangle)\\
&=&\overline{\nabla}^{i}(\langle(\nabla_{i}+A_{i})\psi,\psi\rangle
+\langle\psi,(\nabla_{i}+A_{i})\psi\rangle)\\
&=&\langle\overline{\nabla}^{i}(\nabla_{i}\psi),\psi\rangle+\langle(\overline{\nabla}^{i}A_{i})\psi,\psi\rangle
+\langle A_{i}\overline{\nabla}_{i}\psi,\psi\rangle\\
& &+\langle(\nabla_{i}+A_{i})\psi,\overline{\nabla}_{i}\psi\rangle
+\langle\overline{\nabla}^{i}\psi,(\nabla_{i}+A_{i})\psi\rangle
+\langle\psi,\overline{\nabla}^{i}(\nabla_{i}\psi)\rangle\\
& &+\langle\psi,A_{i}\overline{\nabla}^{i}\psi\rangle
+\langle\psi,(\overline{\nabla}^{i}A_{i})\psi\rangle\\
&=&
\langle\nabla^{i}\nabla_{i}\psi,\psi\rangle+\langle A^{i}\nabla_{i}\psi,\psi\rangle
+\langle(\overline{\nabla}^{i}A_{i})\psi,\psi\rangle\\
& &
+\langle A_{i}\nabla^{i}\psi,\psi\rangle
+\langle A_{i}A^{i}\psi,\psi\rangle+|\nabla\psi|^{2}+\langle\nabla_{i}\psi,A^{i}\psi\rangle+\langle
A_{i}\psi,\nabla^{i}\psi\rangle\\
& &
+|A\psi|^{2}+|\nabla\psi|^{2}+\langle\nabla^{i}\psi,A_{i}\psi\rangle
+\langle A^{i}\psi,\nabla_{i}\psi\rangle+|A\psi|^{2}\\
& &
+\langle\psi,A^{i}\nabla_{i}\psi\rangle+\langle\psi,A_{i}\nabla^{i}\psi\rangle
+\langle\psi,A^{i}A_{i}\psi\rangle+\langle\psi,(\overline{\nabla}^{i}A_{i})\psi\rangle.
\end{eqnarray*}
Now use the Lichnerowicz formula to obtain
\begin{eqnarray*}
\overline{\Delta}|\psi|^{2}&=&\frac{1}{8}(\langle\mathcal{R}\psi,\psi\rangle
+\langle\psi,\mathcal{R}\psi\rangle)+2|\nabla\psi|^{2}+2|A\psi|^{2}\\
& &+2(\langle A^{i}\nabla_{i}\psi,\psi\rangle+\langle\psi,A^{i}\nabla_{i}\psi\rangle)\\
& &+2(\langle\nabla_{i}\psi,A^{i}\psi\rangle+\langle A_{i}\psi,\nabla^{i}\psi\rangle)\\
& &+\langle(\overline{\nabla}^{i}A_{i})\psi,\psi\rangle+\langle\psi,(\overline{\nabla}^{i}A_{i})\psi\rangle\\
& &+\langle A_{i}A^{i}\psi,\psi\rangle+\langle \psi,A_{i}A^{i}\psi\rangle,
\end{eqnarray*}
where
\begin{equation*}
\mathcal{R}=\overline{R}-2|\overline{E}|^{2}-(\overline{\div}\!\text{ }\overline{E})c(e_{0}).
\end{equation*}

In what follows, for convenience we will neglect the notation $c(e_{i})\psi$
for Clifford multiplication, and instead simply write $e_{i}\psi$. Observe that
\begin{eqnarray*}
\langle \overline{E}e_{i}e_{0}\overline{E}e_{i}e_{0}\psi,\psi\rangle&=&
-\langle e_{i}e_{0}\overline{E}e_{i}e_{0}\psi,\overline{E}\psi\rangle\\
&=&\langle e_{0}\overline{E}e_{i}e_{0}\psi,e_{i}\overline{E}\psi\rangle\\
&=&\langle \overline{E}e_{i}e_{0}\psi,e_{0}e_{i}\overline{E}\psi\rangle\\
&=&-\langle \overline{E}e_{i}e_{0}\psi,e_{i}e_{0}\overline{E}\psi\rangle\\
&=&\langle \overline{E}e_{i}e_{0}\psi,e_{i}\overline{E}e_{0}\psi\rangle\\
&=&\langle \overline{E}e_{i}e_{0}\psi,(-\overline{E}e_{i}-2\overline{g}(\overline{E},e_{i}))e_{0}\psi\rangle\\
&=&-\langle \overline{E}e_{i}e_{0},\overline{E}e_{i}e_{0}\psi\rangle
-2\overline{g}(\overline{E},e_{i})\langle \overline{E}e_{i}e_{0}\psi,e_{0}\psi\rangle.
\end{eqnarray*}
Here we have used the basic law of Clifford multiplication, namely
\begin{equation*}
c(X)c(Y)+c(Y)c(X)=-2\widetilde{g}(X,Y),
\end{equation*}
where $\widetilde{g}$ denotes the spacetime metric. Furthermore, we also observe that
\begin{eqnarray*}
\langle \overline{E}e_{i}e_{0}\psi,e_{0}\psi\rangle&=&\langle e_{0}\overline{E}e_{i}e_{0}\psi,\psi\rangle\\
&=&-\langle \overline{E}e_{0}e_{i}e_{0}\psi,\psi\rangle\\
&=&\langle \overline{E}e_{i}e_{0}e_{0}\psi,\psi\rangle\\
&=&\langle \overline{E}e_{i}\psi,\psi\rangle,
\end{eqnarray*}
and therefore
\begin{equation*}
\overline{g}(\overline{E},e_{i})\langle \overline{E}e_{i}e_{0}\psi,e_{0}\psi\rangle=
\overline{g}(\overline{E},e_{i})\langle \overline{E}e_{i}\psi,\psi\rangle=
\langle \overline{E}\!\text{ }\overline{E}\psi,\psi\rangle=|\overline{E}|^{2}|\psi|^{2}.
\end{equation*}
It follows that
\begin{equation*}
\langle A_{i}A^{i}\psi,\psi\rangle=\langle\psi,A_{i}A^{i}\psi\rangle
=-|A\psi|^{2}+\frac{1}{2}|\overline{E}|^{2}|\psi|^{2}.
\end{equation*}
Moreover
\begin{equation*}
\overline{\nabla}_{i}A_{j}=\frac{1}{2}(\overline{\nabla}_{i}\overline{E})e_{j}e_{0}
+\frac{1}{2}\overline{E}(\overline{\nabla}_{i}e_{j})e_{0}
\end{equation*}
since $\overline{\nabla}_{i}e_{0}=0$, as the Jang initial data is assumed to have no extrinsic curvature.
Also
\begin{eqnarray*}
\langle(\overline{\nabla}_{i}\overline{E})e_{j}e_{0}\psi,\psi\rangle&=&
-\langle\psi,(\overline{\nabla}_{i}\overline{E})e_{j}e_{0}\psi\rangle
-2\overline{g}(\overline{\nabla}_{i}\overline{E},e_{j})\langle\psi,e_{0}\psi\rangle\\
&=&-\langle\psi,(\overline{\nabla}_{i}\overline{E})e_{j}e_{0}\psi\rangle
-2(\overline{\nabla}_{i}\overline{E}_{j}-\overline{g}(\overline{E},\overline{\nabla}_{i}e_{j}))\langle\psi,e_{0}
\psi\rangle,
\end{eqnarray*}
and
\begin{equation*}
\langle \overline{E}(\overline{\nabla}_{i}e_{j})e_{0}\psi,\psi\rangle
=-\langle\psi,\overline{E}(\overline{\nabla}_{i}e_{j})e_{0}\psi\rangle
-2\overline{g}(\overline{E},\overline{\nabla}_{i}e_{j})\langle\psi,e_{0}\psi\rangle.
\end{equation*}
Hence
\begin{equation*}
\langle(\overline{\nabla}^{i}A_{i})\psi,\psi\rangle
=-\langle\psi,(\overline{\nabla}^{i}A_{i})\psi\rangle
-(\overline{\div}\!\text{ }\overline{E})\langle\psi,e_{0}\psi\rangle.
\end{equation*}
Finally we obtain
\begin{eqnarray*}
\overline{\Delta}|\psi|^{2}&=&\frac{1}{8}(\langle\mathcal{R}\psi,\psi\rangle
+\langle\psi,\mathcal{R}\psi\rangle)\\
& &+|\overline{E}|^{2}|\psi|^{2}+2|\nabla\psi|^{2}-(\overline{\div}\!\text{ }\overline{E})\langle\psi,e_{0}\psi\rangle\\
& &+2(\langle A^{i}\nabla_{i}\psi,\psi\rangle+\langle\psi,A^{i}\nabla_{i}\psi\rangle)\\
& &+2(\langle\nabla_{i}\psi,A^{i}\psi\rangle+\langle A_{i}\psi,\nabla^{i}\psi\rangle).
\end{eqnarray*}
It follows that $\phi$ satisfies equation~\eqref{phi:rhs} with a right-hand side given by
\begin{eqnarray*}
\mathcal{F}&=&\frac{1}{8}(\langle\mathcal{R}\psi,\psi\rangle
+\langle\psi,\mathcal{R}\psi\rangle)-\frac{1}{2}\overline{R}|\psi|^{2}\\
& &+|\overline{E}|^{2}|\psi|^{2}+2|\nabla\psi|^{2}-(\overline{\div}\!\text{ }\overline{E})\langle\psi,e_{0}\psi\rangle\\
& &+2(\langle A^{i}\nabla_{i}\psi,\psi\rangle+\langle\psi,A^{i}\nabla_{i}\psi\rangle)\\
& &+2(\langle\nabla_{i}\psi,A^{i}\psi\rangle+\langle A_{i}\psi,\nabla^{i}\psi\rangle).
\end{eqnarray*}

Notice that in the case of equality for the positive mass theorem with charge, $\mathcal{F}=0$ so that~\eqref{phi:rhs}
reduces to the correct
equation for the warping factor $\phi$. Namely, in this case the spinor $\psi$ is covariantly constant with
respect to the connection $\nabla$, $\mathcal{R}=0$, and~\eqref{equality} holds. Thus we obtain
\begin{equation*}
\overline{\Delta}\phi-|\overline{E}|^{2}\phi=0.
\end{equation*}
This is the correct equation for the warping factor of a static spacetime staisfying the Einstein-Maxwell
equations. To see this, observe that if $\widetilde{g}=-\phi^{2}dt^{2}+\overline{g}$ is the static spacetime
metric, then its scalar curvature is given by
\begin{equation*}
\widetilde{R}=\overline{R}-2\phi^{-1}\overline{\Delta}\phi.
\end{equation*}
If $\widetilde{g}$ satisfies the Einstein-Maxwell equations then
\begin{equation*}
\widetilde{R}=-\widetilde{g}^{ab}T_{ab},
\end{equation*}
where $T$ is the stress-energy tensor. Since
\begin{equation*}
T^{ab}=-\left(F^{ac}F_{c}^{b}+\frac{1}{4}\widetilde{g}^{ab}F_{cd}F^{cd}\right)
\end{equation*}
where $F_{ab}$ is the field strength tensor of the electro-magnetic field, we have $\widetilde{R}=0$ so that
\begin{equation*}
\overline{\Delta}\phi-\frac{1}{2}\overline{R}\phi=0.
\end{equation*}
But for (time symmetric) electro-vacuum initial data, we have $\overline{R}=2|\overline{E}|^{2}$, which confirms that
our choice of
warping factor satisfies the correct equation. This is quite amazing, since the choice $\phi=|\psi|^{2}$
was based on an entirely different motivation. Namely, $\phi$ was chosen in order to allow the divergence term
in $\overline{R}$ to be integrated away in the Lichnerowicz formula.



%

\end{document}